\documentclass[preprint,superscriptaddress,showpacs,preprintnumbers,amsmath,amssymb]{revtex4}
\usepackage{graphicx}
\usepackage{dcolumn}
\usepackage{bm}

\begin{document}


\title{Enhanced up-conversion of entangled photons and quantum interference under localized field in nanostructures}

\author{Yoshiki Osaka}
 \email{osaka@pe.osakafu-u.ac.jp}
\author{Nobuhiko Yokoshi}
 \email{yokoshi@pe.osakafu-u.ac.jp}
\author{Masatoshi Nakatani}
\author{Hajime Ishihara}
\affiliation{Osaka Prefecture University, 1-1 Gakuen-cho, Sakai, Osaka 599-8531, Japan}

\date{\today}

\begin{abstract}
We theoretically investigate up-conversion process of entangled two photons on a molecule, which is coupled by a cavity or nanoscale metallic structure. Within one-dimensional input-output theory, the propagators of the photons are derived analytically and the up-conversion probability is calculated numerically.
It is shown that the coupling with the nanostructure clearly enhances the process. We also find that the enhancement becomes further pronounced for some balanced system parameters such as the quantum correlation between photons, radiation decay and coupling between the nanostructure and molecule. The non-monotonic dependencies are reasonably explained in view of quantum interference between the coupled modes of the whole system. 
This result indicates that controlling quantum interference and correlation is crucial for few-photon nonlinearity, and  provides a new guidance to wide variety of fields, e.g., quantum electronics and photochemistry.
\end{abstract}


\maketitle

Nonlinear optical responses have appeared ubiquitously in modern technologies with photophysics, photochemistry, classical and quantum communication~\cite{nonlinear optics}. As the application range is expanded, in addition to high-power optics using lasers, nonlinearity induced by a few photons is attracting increasingly much attention. Such a few-photon optics can lead to various new technologies, e.g., up-conversion for efficient solar cells~\cite{TTA-UC}, visible-to-telecom frequency conversion of single photon~\cite{Single-VTconv} and two-photon gateway for quantum communication~\cite{2photon-Gate}.  On the other hand, the usage of photons with quantum correlation for two-photon processes has attracted much attentions~\cite{TPA1,TPA2,TPA3,TPA4,TPA5}.  Especially, in quantum information technology,  entangled photons are key issues~\cite{NC}. Moreover, the generation efficiency is growing large recently~\cite{PDC,HPS,ELED,QD&I}, and then they are expected to become a new type of luminous source.  Thus, the study of nonlinear responses by the correlated photons is not only interesting in itself, but also can contribute to further development in opto-science and technology.

In order to enhance the nonlinearity of a few photons, it is useful to utilize absorption saturation of discrete levels, e.g., in a molecule and quantum dot~\cite{discrete}. Besides, it is well-known that one can enhance interactions between photons and nanoscale materials, by introducing cavity in which a localized field is generated~\cite{MicroCavity,cavity}. Furthermore, for example, the usage of resonator can make spatial configuration of the order of the nano- or micrometers in electric field intensity~\cite{nanocavity,antinode1,antinode2}. Another approach to strong localized field is to introduce localized surface plasmon resonance (LSPR) in metallic nanostructures such as nanochips and nanorods~\cite{antenna1,antenna2}.  Nearby the structures, localized fields of extraordinary high intensity are generated, and they have steep gradient of nanometer scale in intensity. Such localized fields can break long-wavelength approximation, and are expected to open a new type of optics including dipole-forbidden excitation. Actually, the existence of such forbidden excitations is suggested theoretically~\cite{forbidden,forbidden2} and experimentally~\cite{forbidden3}. Then, by embedding discrete levels in cavity or nearby nano-scale metallic structure, one can prepare a challenging stage for the studies of nonlinear few-photon optics. 

In this work, we focus on two-photon up-conversion process on a molecular complex system (or quantum dot complex system) coupled to localized field in nanostructures. Here we assume that the two photons have correlated in space. Additionally, the localized field in the nanostructure is assumed to have spatial inhomogeneity, and then both dipole-allowed and -forbidden states through the field can be accessed. We use a fairly intuitive model for the system in order to extract the essential effect of quantum mechanical correlations and coherence.  We analytically derive the propagator for the up-conversion, and numerically calculate the wave function of the converted output photon. As a result, we find that the spatial correlation militates well for the process, and that there exists the regime where the photons rarely induces the up-conversion without entanglement. Furthermore, there is some suitable set of system parameters. The origin of the non-monotonic dependencies can be interpreted in view of quantum interference between the coupled modes of the open cavity-molecule system. It holds the line against the rapid intuition that nonlinearities of photons favor the stronger cavity-molecule couplings than the relaxation constant of the cavity. Although there are some studies for  up- and down-conversions in cavity~\cite{CQED-UC1, CQED-UC2}, the viewpoint on the interference between the cavity and molecule is rarely met. Therefore, our result can provide a guideline for nonlinear optical reactions, and contributes to future applications of weak light physics. 

\begin{figure}[p]%
\begin{center}
\includegraphics[width=80mm]{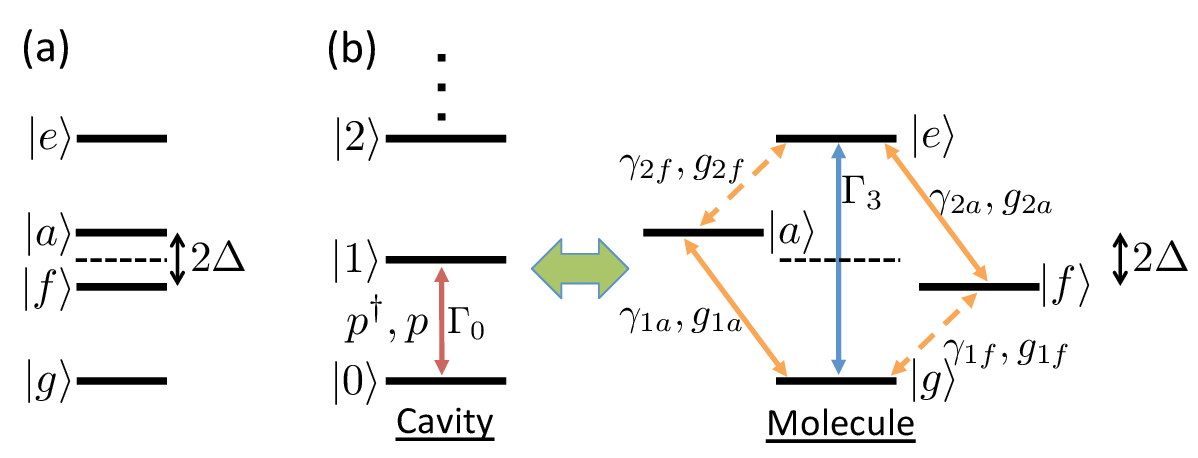}
\end{center}
\caption{(a) Schematic illustration of the levels in the complex molecule (or quantum dot). The energy difference between $|a\rangle$ and $|f\rangle$ is $2\Delta$. (b) Scattering processes in the coupled system. The localized cavity mode with large relaxation constant $\Gamma_0$ excites the four-level molecule. The dipole-allowed excitations occur in the molecule with the constants $\{g_{1a}, g_{2a}\}$, and the dipole-forbidden ones with $\{g_{1f}, g_{2f}\}$. The relaxation constants in the molecule $\{\gamma_{1a}, \gamma_{1f}, \gamma_{2a}, \gamma_{2f}, \Gamma_3\}$ are set to be much smaller compared with $\Gamma_0$.}
\label{model}
\end{figure}
The system under consideration is shown in Fig.~\ref{model}(a). The molecule has four levels; the states $|a\rangle$ and $|e\rangle$ are dipole-allowed from the ground state $|g\rangle$, and the state $|f\rangle$ is dipole-forbidden owing to the parity of the wave function. For example, a phthalocyanine dimer has almost A$_{\rm u}$ symmetry for the highest occupied molecular orbital (HOMO). The lowest unoccupied molecular orbital (LUMO) of each phthalocyanine with B$_{\rm g}$ symmetry is split into two states, i.e., allowed and forbidden states by dipole-dipole interaction~\cite{PcH2}. The excited states, which corresponds to $|e\rangle$, has almost A$_{\rm u}$ symmetry. In addition, for a tetramer, the various allowed and forbidden states exist as shown in supplementary material \cite{supple}.

The total Hamiltonian is $H=H_c+H_m+H_p+H_{cp}+H_{mp}+H_{cm}$. Here we model the cavity mode (or polarization mode by LSPR) by simple boson, i.e., $H_c=\hbar \omega_c p^{\dagger}p$ with $p$ being the annihilation operator. As for the molecule, the Hamiltonian is $H_m=\hbar \omega_a\sigma_{aa}+  \hbar \omega_f\sigma_{ff}+ \hbar \omega_e\sigma_{ee}$, in which $\sigma_{mn} =|m\rangle \langle n|$ with  $\{ m,n \} = \{ g,a,f,e \}$ and the resonant energies are measured from the ground state $|g\rangle$. The resonant energies of $|a\rangle$ and $|f\rangle$ differ from each other by $2\Delta$. In this work, we employ the one-dimensional mode and then the photon field can be described as $H_p=\int dk \hbar ck (\tilde{a}_{k}^{\dagger}\tilde{a}_{k}  + \tilde{b}_{k}^{\dagger}\tilde{b}_{k})$. Here the operator $b_r$ annihilates an up-converted photon at position $r$,  and  $a_r$ annihilates an input or unconverted output photon. The tilde on the operators indicates Fourier transformation, e.g., $\tilde{a}_k=\sqrt{1/2\pi} \int  dk  a_r e^{-ikr}$. Although the model seems oversimplified, most of the systems is expressed by superposing the one-dimensional cases, and then this simplification can reasonably extract an essential aspect of the problem.

Input two photons interact with the coupled system at the origin $(r=0)$. The interactions in the molecule are summarized in Fig.~\ref{model}(b). Then, the cavity-photon coupling is written within rotating-wave approximation as 
\begin{eqnarray}
H_{cp}=i \hbar \sqrt{c\Gamma_0} (  p^{\dagger}a_{r=0} -a_{r=0}^{\dagger} p  ),
\end{eqnarray}
in which $\Gamma_0$ characterize the decay of the cavity mode. In a similar fashion, the radiations of the photons from the molecule are 
\begin{eqnarray}
H_{mp}&=&i \hbar \sqrt{c\gamma_{1_a}}  \sigma_{ag}a_{r=0} 
+i \hbar \sqrt{c\gamma_{2_a}}  \sigma_{ef}a_{r=0}
\nonumber\\
&&
+i \hbar \sqrt{c\gamma_{2_f}}  \sigma_{ea}a_{r=0}
+i \hbar \sqrt{c\gamma_{1_f}}  \sigma_{fg}a_{r=0}
\nonumber\\&&
+i \hbar \sqrt{c\Gamma_3}   \sigma_{eg}b_{r=0} 
+ \rm{h.c.} ,
\end{eqnarray}
with the constants  $\{\gamma_{1a}, \gamma_{1f}, \gamma_{2a}, \gamma_{2f}, \Gamma_3\}$. In general, the relaxation constants of the forbidden transitions are much smaller than the ones of the allowed transitions ($\gamma_{ia} \gg \gamma_{if}$). The last part of the Hamiltonian is for the cavity-molecule couplings. Here the localized field with spatial gradient is assumed to produce similar intensities of absorptions in both the dipole-allowed and -forbidden transitions~\cite{ET}. Then, in the Hamiltonian
\begin{eqnarray}
H_{cm}&=&
\hbar g_{1_a}  \sigma_{ag}p +\hbar g_{2_f}\sigma_{ea} p
\nonumber \\
&&
+\hbar g_{1_f} \sigma_{fg}p+\hbar g_{2_a}\sigma_{ef} p+\rm{h.c.},
\end{eqnarray}
the coupling constants $\{g_{1a}, g_{1f}, g_{2a}, g_{2f}\}$ is considered to have the magnitudes of the same order.

We analyze the whole process by using one-dimensional input-output theory.  As an initial state, we prepare the input state vector which can be written as
\begin{eqnarray}
|\psi_{\rm in}\rangle=\int dr_1 dr_2 ~ \frac{f(r_1,r_2)}{\sqrt{2}}a_{r_1}^{\dagger}a_{r_2}^{\dagger}|V\rangle.
\end{eqnarray}
Here $|V\rangle$ represents the vacuum state of the whole system, i.e., $|0\rangle \otimes |g\rangle$ with zero photon. The function $f(r_1,r_2)$ is the symmetrized two-photon wave function. The theory does not depend on the explicit form of $f(r_1,r_2)$, and then we can take it in arbitrary form. Here we have spatially correlated state as the input expressed by bi-variable Gaussian pulse as
\begin{eqnarray}
f(r_{1},r_{2})
&=&
\frac{\exp 
\left[ -\frac{\bar{r}_1^2+\bar{r}_2^2-2\rho \bar{r}_1 \bar{r}_2}{4(1-\rho^2)d^2} 
+i \frac{\omega_0}{c}(\bar{r}_1+\bar{r}_2)
\right] }
{(2 \pi )^{1/2}d(1-\rho^2)^{1/4}},
\end{eqnarray}
where $\bar{r}=r-a$ is the distance from the initial position $a$, and $d$ is the pulse length. The frequency of the pulse $\omega_0$ is set to be resonant to the $\omega_c$. When the correlation parameter $\rho$ is equal to 0, the input two photons can be decoupled. On the other hand, as $\rho$ gets close to $1$ $(-1)$, they are in strongly correlated (anti-correlated) state. Such an entangled photon-pair can be actually generated using spontaneous parametric down-conversion~\cite{QD&I}.

The output state becomes superposition of the two vectors; one is the up-converted photon state, and the other is the two-photon state as in the input. Here we define the wave functions of the two vectors respectively as
\begin{eqnarray}
h(r;\tau)=\int dr_1' dr_2' G_1(r,r_1',r_2';\tau)f(r_1',r_2'),~~~~~~~\label{h.eq}
\\
g(r_1,r_2;\tau)=\int dr_1' dr_2' G_2(r_1,r_2,r_1',r_2';\tau)f(r_1',r_2'),\label{g.eq}
\end{eqnarray}
where $G_1(r,r_1',r_2';\tau)$ ($G_2(r_1,r_2,r_1',r_2';\tau)$) represents the propagator for the up-conversion process (two-photon emission one). Here the time $\tau$ is taken to be sufficiently far from the interaction point to the coupled system. 

In calculating the propagators as a practical matter,  we employ the method developed in Ref.~\cite{method}, in which a coherent state $|\phi \rangle$ of $a_r$ is introduced. Then, one finds that $a_r| \phi \rangle= \sum_{j=1,2} \mu_j \delta(r-r_j')| \phi \rangle$ and $b_r| \phi \rangle=0$, where $\{\mu_j\}$ are perturbation coefficients. From the Heisenberg equations for the operators, we obtain the simultaneous equations for the expectation values of them. Within the first order of $\mu_i$ and $\mu_1\mu_2$, it is found that, e.g.,
\begin{eqnarray}
G_1(r,r_1,r_2;\tau)&\equiv& \frac{1}{\sqrt{2}} \langle V|  b_r(\tau) a_{r_1}^{\dagger}(0)a^{\dagger}_{r_2}(0)|V \rangle
\nonumber \\
&\propto& \langle \sigma_{ge}(\tau-\frac{r}{c}) \rangle^{\mu_1\mu_2},
\end{eqnarray}
in which $\langle \sigma_{ge}(\tau) \rangle^{\mu_1\mu_2}$ means the perturbation component proportional to $\mu_1\mu_2$ in $\langle \sigma_{ge}(\tau) \rangle$. We can analytically solve the equation, and obtain the propagators and the up-conversion probability $P=\int dr |h(r;\tau)|^2$. The details of the calculation is in supplementary material~\cite{supple}.

Hereafter we assume that the frequency of the cavity mode is set to be $\omega_c=(\omega_a+\omega_f)/2=\omega_e/2$. Then, the up-converted photon has twice the frequency of that of the input photon $\omega_0$.  Because the relaxation constant of the cavity mode is large compared to the other rates, we use $\Gamma_3/\Gamma_0=0.2$, $\gamma_{1a,2a}/\Gamma_0=0.01$, $\gamma_{1f,2f}/\Gamma_0=0.001$. In addition, for simplicity, we assume that all the cavity-molecule coupling constants are equal, i.e.,  $g_{1a}=g_{1f}=g_{2a}=g_{2f}=g$. 

\begin{figure}[p]%
\begin{center}
\includegraphics[width=70mm]{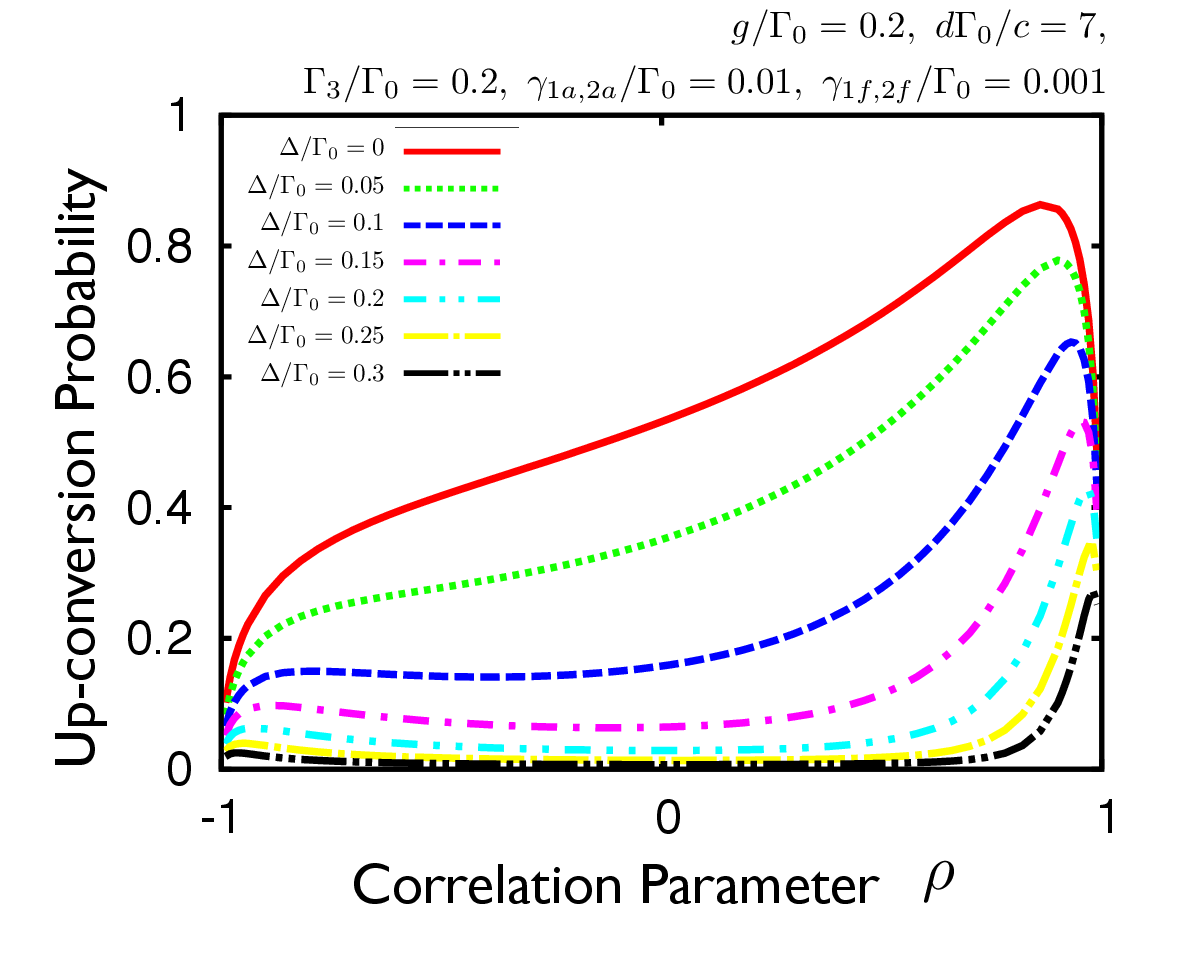}%
\end{center}
  \caption{%
Up-conversion probabilities are plotted against correlation parameter $\rho$. Here we employ $g/\Gamma_0=0.2$ and $d\Gamma_0/c=7$. We set the parameters to be $\Gamma_3/\Gamma_0=0.2$, $\gamma_{1a,2a}/\Gamma_0=0.01$, $\gamma_{1f,2f}/\Gamma_0=0.001$, which are the same also in following figures. One can see that the probabilities take maximum values when $\rho$ is close to 1. As the detuning $\Delta$ increases, the probability decreases. When $\Delta \gtrsim  \Gamma/4$, the up-conversion does not occur except $\rho\sim 1$.
}
\label{delta}
\end{figure}

Figure \ref{delta} shows the correlation parameter dependence of the up-conversion probability for $g/\Gamma_0=0.2$. Here we set the pulse length to be $d\Gamma_0/c=7$, which corresponds to $d=138\mu{\rm m}$ for $\Gamma_0=20{\rm meV}$. The different lines correspond to the ones for different detunings $2\Delta$ between the states $|a\rangle$ and $|f\rangle$. Because the probability for $g=0$ is at most $P \lesssim 0.01$ in the same condition, it is apparent that the couplings to the cavity enhance the up-conversion. We calculate $\rho$-dependence of  the up-conversion probability, and find that it  exceeds $P = 0.8$ when the input two-photon is correlated state ($\rho = 0.9$). 
The enhancement is a similar effect of the increased nonlinearities reported in Refs. \cite{TPA3, TPA4, TPA5}. On the other hand, as the detuning increases, the probability decreases. When the detuning $2\Delta$ becomes larger than the broadening due to the  cavity-radiation coupling ($\sim \Gamma_0/2$), the probability is negligibly small for non-correlated photon-pair ($\rho=0$).  However, as the correlation becomes strong, it exceeds $P=0.4$ even when $\Delta/\Gamma_0=0.3$. Just near $\rho=1$, the up-conversion probability turns to decrease. This is because, if the two photons interact with the molecule at the same moment, the sequential process of the up-conversion is inhibited~\cite{upcon}. Moreover, it should be remarked that the input of the entangled photons makes the up-conversion highly robust also against the deviations of the frequencies $\{\omega_0$, $\omega_c$, $(\omega_a+\omega_f)/2 \}$ from the resonance condition, and of the cavity-molecule coupling constants from $\{g_{1a}=g_{1f}=g_{2a}=g_{2f} \}$~\cite{supple}.

\begin{figure}[p]%
\begin{center}
\includegraphics[width=80mm]{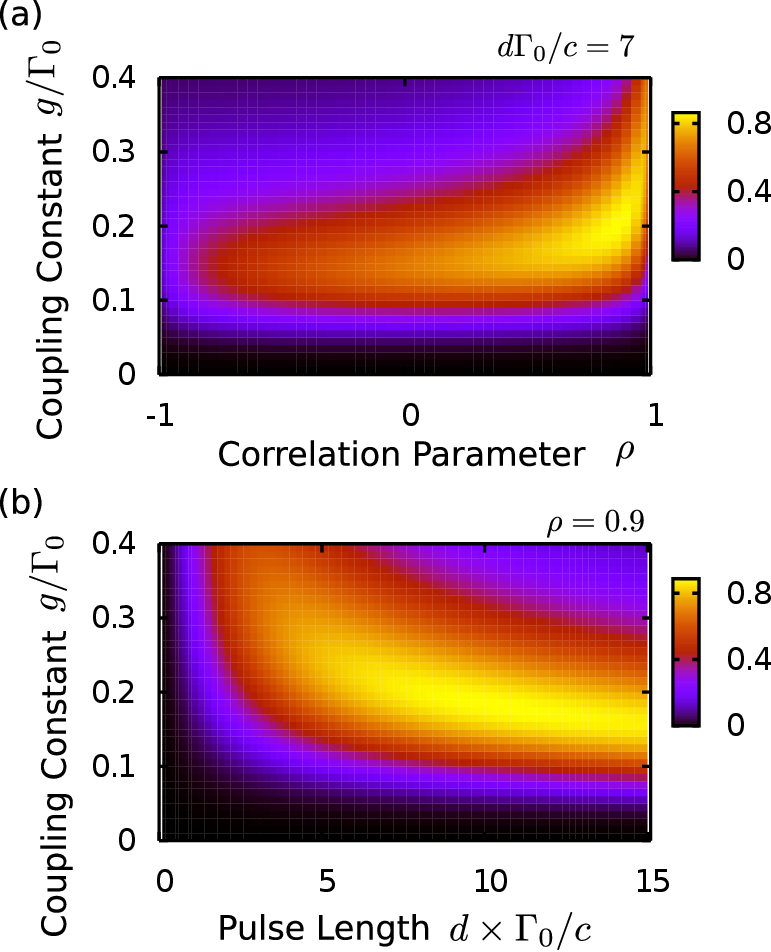}
\end{center}
\caption{%
(a) Up-conversion probability is plotted against the coupling constant $g$ and the correlation parameter $\rho$. Here we set the detuning and pulse length to $\Delta=0$ and $d\Gamma_0/c=7$. It is apparent that the large correlation ($\rho \sim 1$) is preferred for the up-conversion.  (b) The plot against $g$ and $d$ for $\Delta=0$ and $\rho=0.9$, which shows the long pulse has advantage for the conversion. In both the figures, one can see that the optimal regime for $g$ exists between 0.15 and 0.2. 
}
\label{rho-g}
\end{figure}

Subsequently, we investigate the dependence on the cavity-molecule coupling. In Fig.~\ref{rho-g}(a), we plot the up-conversion probability as the function of  the correlation parameter $\rho$  and coupling $g$ for $\Delta=0$. As $g/\Gamma_0$ increases from zero, due to Purcell effect~\cite{effective-coupling}, the probability becomes large by  the increase of the effective photon-molecule coupling $\sim 4g^2/\Gamma_0$ in the weak coupling regime ($g/\Gamma_0 < 0.1$)~\cite{supple}. It should be noted that the optimum parameter regime exists, where $g/\Gamma_0=0.15 \sim 0.2$. This can be explained in view of quantum mechanical interference between the coupled modes in the open system. When we focus on the first excitation by one of the photons, the system can be seen as V-type three level coupled to the cavity. Then, three eigenmodes can be considered; one dark mode and two bright modes. Neglecting the radiation decay in the molecule, the effective bright modes in the open system can be simplified as
\begin{eqnarray}
E_{\pm}=\frac{\hbar}{2}\Bigl( \omega_a+\omega_c-i\frac{\Gamma_0}{2} \pm \sqrt{(\omega_a-\omega_c+i\frac{\Gamma_0}{2})^2+8g^2} \Bigr),
\end{eqnarray}
for $\Delta=0$. When $\Gamma_0=0$, the level splitting between the two bright modes becomes $2\sqrt{2}g$ at the anti-crossing point. Then the splitting of three eigenmodes is disappeared in the presence of the broadening by the cavity for $\Gamma_0/2 \geq 2\sqrt{2}g$, i.e., they oscillate effectively in the same frequency~\cite{supple}. This is the same physics as the damped Rabi oscillation in open quantum mechanics~\cite{rabi}. Thus the two bright modes, both the cavity mode and molecule level are included, can interfere constructively and destructively. When the destructive interference occurs in the cavity, it is possible to make only the molecule oscillate, i.e., the ground state of the cavity becomes transparent for the incident photon. 

In addition, sufficient interference also requires the two coupled modes to decay with the same rate. Such a situation is achieved when the two modes are well-superposed by the cavity-molecule coupling. Indeed, the imaginary parts of the eigenmodes are found to coincide at $\omega_c=\omega_a$ for  $\Gamma_0/2 \leq 2\sqrt{2}g$~\cite{supple}. Almost the same circumstance exists for the second excitation to the state $|e\rangle$; the lead difference comes from the radiation of the up-converted photon by $\Gamma_3$.  Therefore, the up-conversion is considered to be enhanced when $\Gamma_0/2 \sim 2\sqrt{2}g$, which is consistent with the result in Fig.~\ref{rho-g}(a).

In Fig.~\ref{rho-g}(b), we show the conversion probability against the pulse length and cavity-molecule coupling for strongly correlated photons ($\rho=0.9$). One can see that the long pulse length has advantage for the quantum interference discussed above. This is because, in the long pulse, the frequency component with $\omega\not= \omega_0$ does less contributes to the system.  Indeed, assuming the usage of weak monochromatic light, similar discussion for the effective excitation is presented within linear response theory, where a molecule is assumed to be embedded near the metallic structure~\cite{ET}. Recent experiment presented the result closely related to such a phenomenon using the nanostructure of Au~\cite{metal-gap}. Besides, with the use of relatively narrow-band laser pulses, similar phenomenon was reported for third-harmonics generation by hybrid plasmonic-dielectric compound~\cite{nonlinear ET}. Then, using the entangled pair of long pulse, further application of few-photon nonlinearity can be expected. 

In conclusion, we have analyzed up-conversion process of spatially entangled two photons on a molecule, which is coupled by a cavity or nanoscale metallic structure. As a result, we found that the usage of the entangled pair makes the up-conversion further facile. We also elucidated suitable conditions and their origins for the input photons to efficiently excite the molecule. This phenomenon is caused by quantum interference between two eigenmodes, which include cavity mode and molecule. Thus, the results imply that controlling the quantum correlation and interference leads to high-efficient nonlinear few-photon optics.

Our fairly simple model enabled us to discuss about complicated four-body (two photons, cavity and molecule) interaction and enriched our understanding about underlying physics and its dependence on key parameters in cavity-molecule coupled system. However, of course, it is quite important for practical applications to take into account individual circumstances of particular systems. Then, we evaluated the cavity-molecule couplings for a specific gold nanostructure as the antenna system, and showed that our findings can be verified in realistic nanostructures~\cite{supple}. In the demonstration for the large couplings $g_{1a,f}$ within the linear response, we also have taken into account quenching effects of the molecules due to the excitation energy transfer from the molecule to the antenna~\cite{quenching1, quenching2}. Few-photon responses have recently appeared not only in quantum physics but also in photochemistry, e.g., two-photon photopolymerization in SU-8 molecules embedded on Au nanostructure~\cite{2photon}. Thus presenting an intuitive guideline for efficient nonlinear responses can contribute also to further promoting more elaborate studies in various fields, and development in future photo-science. 

We thank Dr. Y. Mizumoto for helpful discussion on numerical calculation in supplementary material. This work was partially supported by Grant-in-Aid for Scientific Research (KAKENHI) No.~19049014 on Priority Area ``Strong Photons-Molecules Coupling Fields 470'', a Grant-in-Aid for Challenging Exploratory Research No.~23654105, and a Grant-in-Aid for Young Scientists (B) No.~24760046 from MEXT, Japan.

\end{document}